\documentclass[%
 aip,
 rsi,%
 amsmath,amssymb,
 reprint,%
]{revtex4-1}
\pdfoutput=1
\usepackage{graphicx}% Include figure files
\usepackage{dcolumn}% Align table columns on decimal point
\usepackage{bm}
\usepackage{dcolumn}% Align table columns on decimal point
\usepackage{color}
\usepackage{lipsum}
\usepackage{CJKfntef}
\usepackage{mathrsfs}
\usepackage{enumitem}
\usepackage{slashed}
\usepackage{graphicx}
\usepackage{amsmath,amssymb,amsfonts}
\usepackage{hyperref}
\usepackage{pifont}
\usepackage[perpage]{footmisc}
\allowdisplaybreaks[1]

\usepackage{tikz}
\usetikzlibrary{trees}
\usetikzlibrary{decorations.pathmorphing}
\usetikzlibrary{decorations.markings}

\definecolor{jblue}  {RGB}{20,50,100}
  \definecolor{npurple}  {RGB} {153, 51, 204}
  \definecolor{wred}   {RGB}{217,0,56}
  \definecolor{white}   {RGB}{255,255,255}

  \definecolor{korange}   {RGB}{235, 80,  43}
  \definecolor{korange2}   {RGB}{245, 100,  63}
  \definecolor{kyelloworange}   {RGB}{255, 210,  110}
  \definecolor{kyelloworange2}   {RGB}{240, 170,  90}
  \definecolor{kred}   {RGB}{204,  102, 153}
  \definecolor{kpurple}   {RGB}{153,  61, 190}
  \definecolor{kpurplelight}   {RGB}{213,  161, 230}

\tikzset{
	  photon/.style={decorate, decoration={snake}, draw=npurple,very thick},
	  boson/.style={decorate, decoration={snake}, draw=npurple,very thick},
	  electron/.style={draw=jblue,very thick, postaction={decorate},
	           decoration={markings,mark=at position .55 with {\arrow[draw=jblue]{>}}}
	  },
	  electron2/.style={draw=jblue,very thick, postaction={decorate},
	           decoration={markings,mark=at position .55 with {\arrow[draw=jblue]{<}}}
	  },
	  fermion/.style={draw=jblue,very thick, postaction={decorate},
	            decoration={markings,mark=at position .55 with {\arrow[draw=jblue]{}}}
	  },
	  gluon/.style={decorate, draw=korange,very thick, %kred
	    decoration={coil,amplitude=4pt, segment length=6pt}},
	  higgs/.style={draw=wred,very thick, postaction={decorate},
	           decoration={markings,mark=at position .55 with {\arrow[draw=wred]{>}}}
	  },
	  nothing/.style={draw=white,very thick}
	}

\tikzset{
	    vector/.style={decorate, decoration={snake}, draw},
	provector/.style={decorate, decoration={snake,amplitude=2.5pt}, draw},
	antivector/.style={decorate, decoration={snake,amplitude=-2.5pt}, draw},
    fermion/.style={draw=black, postaction={decorate},
       decoration={markings,mark=at position .55 with {\arrow[draw=black]{>}}}},
   fermionbar/.style={draw=black, postaction={decorate},
       decoration={markings,mark=at position .55 with {\arrow[draw=black]{<}}}},
    fermionnoarrow/.style={draw=black},
   gluon/.style={decorate, draw=black,
        decoration={coil,amplitude=4pt, segment length=5pt}},
    scalar/.style={dashed,draw=black, postaction={decorate},
        decoration={markings,mark=at position .55 with {\arrow[draw=black]{>}}}},
    scalarbar/.style={dashed,draw=black, postaction={decorate},
        decoration={markings,mark=at position .55 with {\arrow[draw=black]{<}}}},
    scalarnoarrow/.style={dashed,draw=black},
    electron/.style={draw=black, postaction={decorate},
        decoration={markings,mark=at position .55 with {\arrow[draw=black]{>}}}},
	 bigvector/.style={decorate, decoration={snake,amplitude=4pt}, draw},
}

\begin{document}

\title{Dark photon production via $\gamma \gamma \rightarrow \gamma A'$}

\author{Xiaorui Wong}
\email{xiaorui$\_$wong@pku.edu.cn}
\affiliation{School of Physics, Peking University, Beijing 100871, China}
\affiliation{Institute of High Energy Physics, Chinese Academy of Sciences, Beijing 100049, China }

\author{Yongsheng Huang}
\email{huangys82@ihep.ac.cn}
\thanks{Corresponding author}
\affiliation{Institute of High Energy Physics, Chinese Academy of Sciences, Beijing 100049, China }

\date{\today}

\begin{abstract}
The dark photon is a new gauge boson which arises from an extra $U'(1)$ gauge symmetry.
In this paper, a novel dark photon production mechanism based on MeV-scale $\gamma$-$\gamma$ collider is considered: $\gamma \gamma \rightarrow \gamma A'$.
With the aid of PACKAGE-X, differential cross section of $\gamma \gamma \rightarrow \gamma A'$ is obtained, as a function of the kinetic mixing parameter $\varepsilon$ and dark photon mass $m_{A'}$. Taking the light-by-light scattering as background, the constraints on the dark photon parameter space for different time intervals in a MeV-scale $\gamma$-$\gamma$ collider are also given.
\end{abstract}

\maketitle
\flushbottom

\section{Introduction}
The dark matter problem has never lost its allure for nearly a century\cite{ref1.1.1, ref1.1.2,ref1.1.3,ref1.1.4}.
Back to 1922, Jacobus Kapteyn first suggested the existence of dark matter by stellar velocities~~\cite{ref1.1.5}. In 1933, Fritz Zwicky postulated dark matter by the huge discrepancies between luminous mass and dynamical mass of the Coma Cluster~\cite{ref1.1.6}.
There are many observational evidences for dark matter in cosmology: galaxy rotation curve~\cite{ref1.1.7,ref1.1.8,ref1.1.9}, gravitational lensing~\cite{ref1.1.10,ref1.1.11}, bullet cluster~\cite{ref1.1.12}, cosmic microwave background~\cite{ref1.1.13,ref1.1.14} and so on.

Dark matter candidates include: weakly interacting massive particles(WIMPs), asymmetric dark matter, axions, sterile neutrinos, dark photon, etc.
WIMPs are expected to have been thermally produced in the early universe, they have large mass compared to standard particles, and they interact with cross sections no higher than the weak scale~\cite{ref1.1.16}.
Models of asymmetric dark matter are based on the idea that dark matter may carry a matter-antimatter asymmetry~\cite{ref1.1.19}.
The existence of axions was first postulated to solve the strong CP problem of quantum chromodynamics(QCD)~\cite{ref1.1.20}.
Sterile neutrinos are neutrinos with right-handed chirality, and they interact only via gravity~\cite{ref1.1.23}.
~\\

Dark photon is the gauge boson of an extra $U'(1)$ symmetry~\cite{ref1.1.24}, and can interact with Standard Model photon via kinetic mixing~\cite{ref1.1.25,ref1.1.26,ref1.1.27}.
Dark photon could be either massless or massive in different frames. The massive case has gained a lot of attention because it couples directly to the standard model currents and is more readily accessible in experiment~\cite{ref1.1.28}. In this work, massive dark photon is considered.

The production mechanisms of dark photon mainly include: bremsstrahlung~\cite{pb.1}, annihilation ~\cite{pa.1,pa.2,pa.3}, meson decay~\cite{pm.1} and Drell-Yan~\cite{pd.1}.
Many related experiments ~\cite{pb.e1,pb.e2,pb.e3,pb.e4,pb.e6,pb.e7,pa.e1,pa.e2,pa.e3,pm.e1,pm.e2,pd.e1,pd.e2} have been taken. But still, no robust signature of dark photon has come out.

~\\

In this paper, a novel dark photon production mechanism based on $\gamma$-$\gamma$ collider is considered: $\gamma \gamma \rightarrow \gamma A'$. Where $\gamma $ means standard model photon and $A'$ stands for dark photon.
$\gamma$-$\gamma$ collider was firstly suggested in the early 1980s, with a concept of creating it by uniting linear electron accelerators with high peak and average power lasers~\cite{ref3.1}.
There is still no $\gamma$-$\gamma$ colliders in the world, but the technology of building a MeV-scale $\gamma$-$\gamma$ collider is being mature.
Futhermore, the background is respectively clean, which could serve better environment in searching dark photon.
In this paper, differential cross section of the process $\gamma \gamma \rightarrow \gamma A'$ is calculated and limits on dark photon parameter space in a $\gamma$-$\gamma$ colllider are also given.

This paper is organized as follows: in section \ref{mainpart1}, the dark photon model will be given, and amplitude of $\gamma \gamma \rightarrow \gamma A'$ is calculated. In section \ref{mainpart2}, we analyse our results and gave limits on dark photon in a MeV-scale $\gamma$-$\gamma$ collider. Finally, summary is made in section \ref{summary}.

\section{Amplitude of $\gamma \gamma \rightarrow \gamma A'$}\label{mainpart1}

If the Standard Model gauge group is extended by adding a new abelian $U'(1)$ symmetry: $SU(3) \times SU(2) \times U(1) \times U'(1)$, a corresponding new gauge vector boson will occur, which is called dark photon and often labeled by $A'$ or $\gamma'$. With two $U(1)s$ in the gauge group, there will be kinetic mixing, the mixing term can be written as~\cite{ref1.1.26}:
\begin{align}
\mathcal{L}_{mix} = - \dfrac{\varepsilon}{2} F_{\mu\nu} F^{\prime \mu\nu}
\end{align}
The Lagrangian is:
\begin{align}
\mathcal{L} = -\dfrac{1}{4} F_{\mu\nu} F^{\mu\nu} -\dfrac{1}{4} F'_{\mu\nu} F'^{\mu\nu} - \dfrac{\varepsilon}{2} F_{\mu\nu} F^{\prime \mu\nu} + \dfrac{1}{2} m_{A'}^2 V_\mu V^\mu
\end{align}
where $\varepsilon$ is kinetic mixing parameter, $m_{A'}$ is mass of dark photon, $\varepsilon$ and $m_{A'}$ are the only two free parameters. $F_{\mu\nu}$ is the field strength tensor of $U(1)$ and $F'_{\mu\nu} = \partial_\mu V_\nu - \partial_\nu V_\mu$ is the field strength tensor of $U'(1)$.
~\\

Dark photon production is considered via $\gamma \gamma \rightarrow \gamma A'$, this reaction can be taken as the result of the production of a virtual electron-positron pair by two initial photons, followed by annihilation of the pair into the final state photon, and the dark photon.

Define $k_1$ and $k_2$ are momentums of incoming photons, $k_3$ is momentum of outgoing photon and $k_4$ is momentum of outgoing dark photon:
\begin{align}
k_1 &= (\omega , 0 , 0 , \omega) \notag \\
k_2 &= (\omega , 0 , 0 , -\omega) \notag \\
k_3 &= (|\textit{\textbf{k}}_3| , |\textit{\textbf{k}}_3|\sin\theta , 0 , |\textit{\textbf{k}}_3|\cos\theta) \notag \\
k_4 &= (\sqrt{m_{A'}^2 + |\textit{\textbf{k}}_3|^2 } , -|\textit{\textbf{k}}_3|\sin\theta , 0 , -|\textit{\textbf{k}}_3|\cos\theta) \notag
\end{align}
$\hbar \omega$ is energy of an incoming photon, here we set $\hbar = 1$. $\theta$ is the angle between directions of outgoing and incoming photons.

We define Mandelstam variable:
\begin{align}
s = (k_1 + k_2 )^2 , t = (k_1 - k_3 )^2 , u = (k_1 - k_4 )^2
\end{align}
which satisfies:
\begin{align}
s + t + u = m_{A'}^2
\end{align}

Differential cross section of this process is:
\begin{align}\label{formulea}
\dfrac{d \sigma}{d \Omega} = \dfrac{1}{64 \pi^2 (4\omega)^2} \dfrac{|\textit{\textbf{k}}_3|}{|\textit{\textbf{k}}_1|} |\mathcal{M}_{\mathrm{fi}}|^2
\end{align}

Because $k_1 + k_2 = k_3 + k_4$, it is easy to get $\dfrac{|\textit{\textbf{k}}_3|}{|\textit{\textbf{k}}_1|} = 1 -  \dfrac{m_{A'}^2}{s}$.

\begin{figure*}[htbp]
	\begin{center}
\centering
\begin{tikzpicture}[line width=1.5 pt, scale=1]
        %\caption{(a)}
		\draw[vector] (-1,2.5) -- (0,2);
        \node at (-1.2,2.5) {$\gamma$};
		\draw[fermion] (2,2) -- (0,2);
		\draw[vector] (2,2) -- (3,2.5);
        \node at (3.2,2.5) {$\gamma$};
		\draw[vector] (-1,-.5) -- (0,0);
        \node at (-1.2,-.5) {$\gamma$};
		\draw[fermion] (0,0) -- (2,0);
		\draw[vector] (2,0) -- (3,-.5);
        \node at (3.2,-.5) {$A'$};
		\draw[fermion] (0,2) -- (0,0);
		\draw[fermion] (2,0) -- (2,2);
        \draw[->](-.8,2.7)--(-.35,2.45);
        \node at (-.5,2.8) {$k_1$};
        \draw[->] (-.8,-.7) -- (-.35,-.45);
        \node at (-.4,-.8) {$k_2$};
        \draw[->] (2.35,2.45) -- (2.8,2.7);
        \node at (2.45,2.8) {$k_3$};
        \draw[->] (2.35,-.45) -- (2.8,-.7);
        \node at (2.45,-.8) {$k_4$};
        \node at (0,2.2) {$\mu$};
        \node at (2,2.2) {$\nu$};
        \node at (-0.2,.2) {$\rho$};
        \node at (2.3,.2) {$\lambda$};
        \node at (1,2.3) {$p$};
        \node at (-.6,1) {$p+k_1$};
        \node at (1,-.3) {$p+k_1+k_2$};
        \node at (2.7,1) {$p+k_3$};

        \node at (1,-1.5) {(s)};
	
        \begin{scope}[shift={(5,0)}]
%\end{tikzpicture}

%\begin{tikzpicture}[line width=1.5 pt, scale=1]
        %\caption{(b)}
		\draw[vector] (-1,2.5) -- (0,2);
        \node at (-1.2,2.5) {$\gamma$};
		\draw[fermion] (2,2) -- (0,2);
		\draw[vector] (2,2) -- (3,2.5);
        \node at (3.2,2.5) {$\gamma$};
		\draw[vector] (-1,-.5) -- (0,0);
        \node at (-1.2,-.5) {$A'$};
		\draw[fermion] (0,0) -- (2,0);
		\draw[vector] (2,0) -- (3,-.5);
        \node at (3.2,-.5) {$\gamma$};
		\draw[fermion] (0,2) -- (0,0);
		\draw[fermion] (2,0) -- (2,2);
        \draw[->](-.8,2.7)--(-.35,2.45);
        \node at (-.5,2.8) {$k_1$};
        \draw[<-] (-.8,-.7) -- (-.35,-.45);
        \node at (-.4,-.8) {$k_4$};
        \draw[->] (2.35,2.45) -- (2.8,2.7);
        \node at (2.45,2.8) {$k_3$};
        \draw[<-] (2.35,-.45) -- (2.8,-.7);
        \node at (2.45,-.8) {$k_2$};

        \node at (1,-1.5) {(t)};
	 \end{scope}
%\end{tikzpicture}
        \begin{scope}[shift={(10,0)}]
%\begin{tikzpicture}[line width=1.5 pt, scale=1]
        %\caption{(c)}

        \draw[vector] (-1,2.5) -- (0,2);
        \node at (-1.2,2.5) {$\gamma$};
		\draw[fermion] (2,2) -- (0,2);
		\draw[vector] (2,2) -- (3,2.5);
        \node at (3.2,2.5) {$A'$};
		\draw[vector] (-1,-.5) -- (0,0);
        \node at (-1.2,-.5) {$\gamma$};
		\draw[fermion] (0,0) -- (2,0);
		\draw[vector] (2,0) -- (3,-.5);
        \node at (3.2,-.5) {$\gamma$};
		\draw[fermion] (0,2) -- (0,0);
		\draw[fermion] (2,0) -- (2,2);
        \draw[->](-.8,2.7)--(-.35,2.45);
        \node at (-.5,2.8) {$k_1$};
        \draw[->] (-.8,-.7) -- (-.35,-.45);
        \node at (-.4,-.8) {$k_2$};
        \draw[->] (2.35,2.45) -- (2.8,2.7);
        \node at (2.45,2.8) {$k_4$};
        \draw[->] (2.35,-.45) -- (2.8,-.7);
        \node at (2.45,-.8) {$k_3$};

        \node at (1,-1.5) {(u)};
\end{scope}
	\end{tikzpicture}
	\caption{Three Feynman diagrams of $\gamma\gamma \rightarrow \gamma A'$, the other three differ from these only in that the internal electron loop is traversed in the opposite direction. Amplitude of the diagram with clockwise loop current gives the same contribution with anticlockwise current case.}
\label{FD}
\end{center}
\end{figure*}
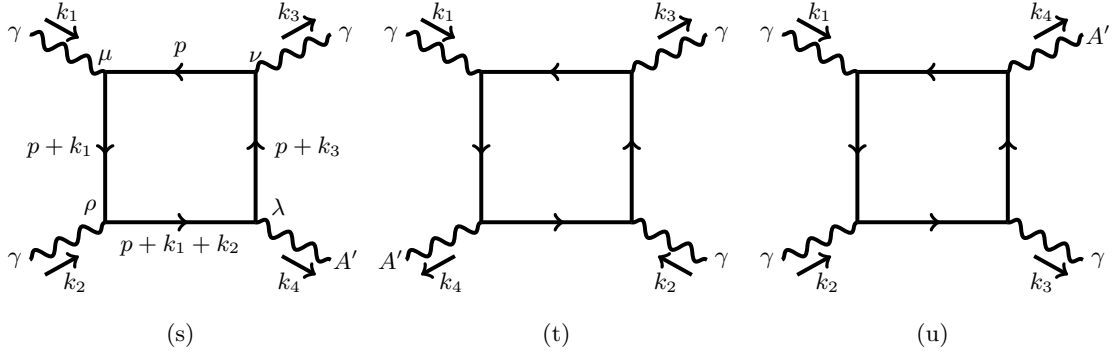

The amplitude of $\gamma \gamma \rightarrow \gamma A'$ has contributions from six diagrams. Three are shown in Figure \ref{FD}, the other three differ from these only in that the internal electron loop traverses in the opposite direction. Amplitude of the diagram with clockwise direction of electrons in the loop is the same as that with the anticlockwise direction loop, so the total scattering amplitude is:
\begin{align}
&\mathrm{i} \mathcal{M}_{\mathrm{fi}}= 2\varepsilon e^4(\Pi_s^{\mu\rho\nu\lambda} + \Pi_t^{\mu\rho\nu\lambda} + \Pi_u^{\mu\rho\nu\lambda}) \epsilon_\mu(k_1) \epsilon_\rho(k_2) \epsilon^{\ast}_\nu(k_3)\epsilon^{\prime\ast}_\lambda(k_4)
\end{align}
where $\epsilon_\mu(k_1) , \epsilon_\rho(k_2)$ are polarisation vectors of incoming photons, $\epsilon^{\ast}_\nu(k_3)$, $\epsilon^{\prime\ast}_\lambda(k_4)$ are polarisation vectors of outgoing photon and dark photon, $ \Pi_s^{\mu\rho\nu\lambda},  \Pi_t^{\mu\rho\nu\lambda},  \Pi_u^{\mu\rho\nu\lambda}$ are loop cotributions of diagrams (s), (t), (u) in  Figure \ref{FD}.

From quantum field theory, we can get:
{\begin{widetext}
\begin{align}
  &\Pi_s^{\mu\rho\nu\lambda}  =  -  \beta^{2\epsilon}\int \dfrac{\mathrm{d}^n p}{(2 \pi)^n}  \dfrac{\mathrm{tr}[\gamma^\mu (\slashed{p} + m)\gamma^\nu (\slashed{p}+\slashed{k_3}+m)\gamma^\lambda (\slashed{p}+\slashed{k_1}+\slashed{k_2}+m ) \gamma^\rho (\slashed{p}+\slashed{k_1}+m)]}{(p^2-m^2+\mathrm{i}0^{+})[(p+k_3)^2-m^2+ \mathrm{i} 0^{+}][(p+k_1+k_2)^2-m^2+\mathrm{i}0^{+}][(p+k_1)^2-m^2+\mathrm{i}0^{+}]
 }
\end{align}
\end{widetext}
}

\noindent where $m$ is electron mass, $\beta$ is an introduced parameter to make the dimension correct, $\epsilon = \dfrac{4-n}{2}$. $ \Pi_t^{\mu\rho\nu\lambda}$ and $ \Pi_u^{\mu\rho\nu\lambda}$ can be got similarly.

Amplitudes are calculated with PACKAGE-X~\cite{ref4.1}. Each diagram in Figure \ref{FD} has divergent contribution, as shown in Eq.(\ref{sdiv}) to (\ref{udiv}) respectively:
\begin{align}\label{sdiv}
\dfrac{4 g^{\lambda \nu} g^{\mu \rho}}{3 \epsilon} + \dfrac{4 g^{\lambda \rho} g^{\mu \nu}}{3 \epsilon} - \dfrac{8 g^{\lambda \mu} g^{\rho \nu}}{3 \epsilon}
\end{align}
\begin{align} \label{tdiv}
 -\dfrac{8 g^{\lambda \nu} g^{\mu \rho}}{3 \epsilon} + \dfrac{4 g^{\lambda \rho} g^{\mu \nu}}{3 \epsilon} + \dfrac{4 g^{\lambda \mu} g^{\rho \nu}}{3 \epsilon}
\end{align}
\begin{align}\label{udiv}
\dfrac{4 g^{\lambda \nu} g^{\mu \rho}}{3 \epsilon} - \dfrac{8 g^{\lambda \rho} g^{\mu \nu}}{3 \epsilon} + \dfrac{4 g^{\lambda \mu} g^{\rho \nu}}{3 \epsilon}
\end{align}
The metric tensor $g^{\mu\nu} = \mathrm{diag}\{1,-1,-1,-1\}$. It is obviously to tell total amplitude has a finite result.

Under low energy limit, taking the $n \rightarrow 4$, the full loop contribution reads as:
\begin{widetext}
\begin{align}
\Pi^{\mu\rho\nu\lambda} = & \Pi_s^{\mu\rho\nu\lambda} + \Pi_t^{\mu\rho\nu\lambda} + \Pi_u^{\mu\rho\nu\lambda} \notag \\ = & \dfrac{1}{m^4} \bigg[ \dfrac{4}{15}k_{1}^\rho k_1^\nu k_2^\mu k_3^\lambda + \dfrac{4}{15} k_{1}^\rho k_2^\mu k_2^\nu k_3^\lambda + \dfrac{4}{15} k_{1}^\rho k_1^\nu k_2^\lambda k_3^\mu - \dfrac{28}{45} k_{1}^\lambda k_1^\rho k_2^\nu k_3^\mu  - \dfrac{28}{45} k_{1}^\rho k_2^\lambda k_2^\nu k_3^\mu  \notag \\ & + \dfrac{28}{45} k_{1}^\rho k_2^\nu k_3^\lambda k_3^\mu - \dfrac{28}{45} k_{1}^\lambda k_1^\nu k_2^\mu k_3^\rho - \dfrac{28}{45} k_{1}^\nu k_2^\lambda k_3^\mu k_3^\rho + \dfrac{4}{15} k_{1}^\lambda k_2^\mu k_2^\nu k_3^\rho + \dfrac{28}{45} k_{1}^\nu k_2^\mu k_3^\lambda k_3^\rho  \notag \\ & - \dfrac{4}{15} k_{1}^\nu k_2^\lambda k_3^\mu k_3^\rho - \dfrac{4}{15} k_{1}^\lambda k_2^\nu k_3^\mu k_3^\rho + \dfrac{28}{45}(m_{A'}^2 - s - t)k_1^\rho k_1^\nu g^{\lambda\mu} - \dfrac{28}{45} t k_1^\rho k_2^\nu g^{\lambda\mu}  \notag \\ & + \dfrac{28}{45} s k_1^\nu k_3^\rho g^{\lambda\mu} - \dfrac{2}{15} (s+t)k_2^\nu k_3^\rho g^{\lambda\mu} - \dfrac{28}{45}(m_{A'}^2 - s - t)k_1^\nu k_2^\mu g^{\lambda\rho}  + \dfrac{28}{45} t k_2^\mu k_2^\nu g^{\lambda\rho}  \notag \\ & - \dfrac{2}{15} (m_{A'}^2 - t)k_1^\nu k_3^\mu g^{\lambda\rho} + \dfrac{28}{45} s k_2^\nu k_3^\mu g^{\lambda\rho}  + \dfrac{2}{15} (m_{A'}^2 - s)k_1^\rho k_2^\mu g^{\lambda\nu} +\dfrac{28}{45} s k_3^\mu k_3^\rho g^{\lambda\nu} \notag \\ & + \dfrac{28}{45} t k_2^\mu k_3^\rho g^{\lambda\nu}  + \dfrac{28}{45}(m_{A'}^2 - s - t)k_1^\rho k_3^\mu g^{\lambda\nu} - \dfrac{14}{45}(m_{A'}^2 - s - t)k_1^\lambda k_1^\nu g^{\mu\rho} \notag \\ &  + \dfrac{14}{45}(m_{A'}^2 - s - t)k_1^\nu k_2^\lambda g^{\mu\rho} + \dfrac{14}{45}t k_1^\lambda k_2^\nu g^{\mu\rho} - \dfrac{14}{45} t k_2^\lambda k_2^\nu g^{\mu\rho}- \dfrac{2}{45}(3s - 7t)k_2^\nu k_3^\lambda g^{\mu\rho} \notag \\ &  + \dfrac{2}{45}(7 m_{A'}^2 - 10s - 7t)k_1^\nu k_3^\lambda g^{\mu\rho}  - \dfrac{14}{45}(m_{A'}^2 - s - t)k_1^\lambda k_1^\rho g^{\mu\nu} - \dfrac{14}{45} s k_1^\lambda k_3^\rho g^{\mu\nu} \notag \\ & + \dfrac{1}{45}(-3m_{A'}^2 s + 3s^2 + 14 m_{A'}^2 t - 14 s t - 14 t^2 )g^{\lambda \nu} g^{\mu \rho}- \dfrac{14}{45}(m_{A'}^2 - s - t) k_1^\rho k_3^\lambda  g^{\mu\nu} \notag \\ & - \dfrac{2}{45}(7 m_{A'}^2 - 7s - 10t)k_1^\rho k_2^\lambda g^{\mu\nu} + \dfrac{1}{45}(14 m_{A'}^2 s -14s^2 - 3 m_{A'}^2 t - 14 s t + 3 t^2 )g^{\lambda \rho} g^{\mu \nu} \notag \\ & + \dfrac{2}{45}( 7s - 3t)k_2^\lambda k_3^\rho g^{\mu\nu} - \dfrac{14}{45} s k_3^\lambda k_3^\rho g^{\mu\nu} - \dfrac{14}{45} t k_2^\lambda k_2^\mu g^{\rho\nu} - \dfrac{14}{45} t k_2^\mu k_3^\lambda g^{\rho\nu} - \dfrac{14}{45} s k_2^\lambda k_3^\mu g^{\rho\nu} \notag \\ & + \dfrac{2}{45}(3 m_{A'}^2 - 3s - 10t)k_1^\lambda k_2^\mu g^{\rho\nu} - \dfrac{2}{45}(3 m_{A'}^2 - 10s - 3t)k_1^\lambda k_3^\mu g^{\rho\nu} - \dfrac{14}{45} s k_3^\lambda k_3^\mu g^{\rho\nu} \notag \\ & + \dfrac{1}{45}(-3m_{A'}^2 s + 3s^2 - 3 m_{A'}^2 t + 20 s t + 3 t^2 )g^{\lambda \mu} g^{\rho \nu}  \bigg]
\end{align}
\end{widetext}

Therefore the amplitude square is:
\begin{align}\label{square}
|\mathcal{M}_{\mathrm{fi}}|^2 = \left(\dfrac{\mathrm{i} \varepsilon e^4}{16 \pi^2} \right)^2 \dfrac{1}{4} |\Pi|^2
\end{align}
where $\dfrac{\mathrm{i}}{16 \pi^2}$ comes from normalization in loop integration, and $1/4$ comes from the average over initial spins.

\begin{figure}
  \centering
  \includegraphics[width=.45\textwidth]{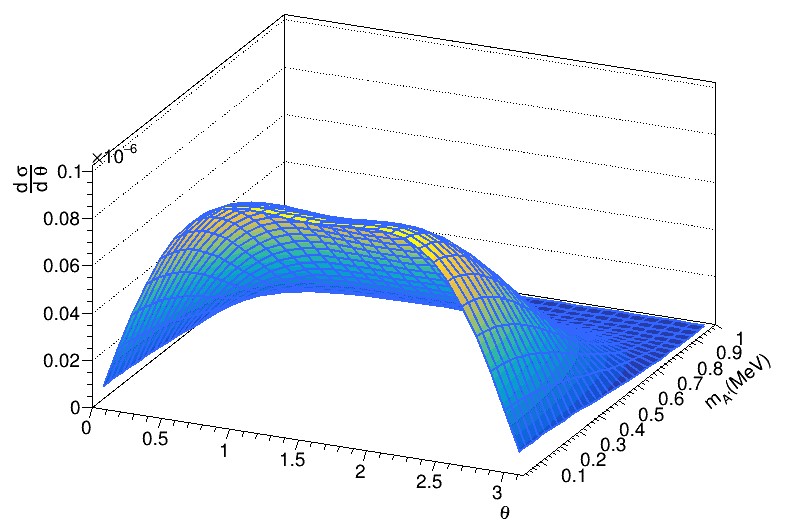}\\
  \caption{Angular distribution of reaction $\gamma \gamma \rightarrow \gamma A'$ (take $\varepsilon = 1$). $\theta$ is the angle between directions of outgoing and incoming photons, and dark photon is considered with mass range $m_{A'} < 1$MeV.}\label{angulardis}
\end{figure}

Take $e = \sqrt{\alpha 4 \pi}$ , $t = \dfrac{s - m_{A'}^2}{2}(\cos\theta -1)$, Eq.(\ref{square}) and $\alpha^2 = r_e^2 m^2$ into Eq.(\ref{formulea}), differential cross section becomes:
\begin{align}\label{difsection}
 \dfrac{\rm{d}\sigma}{\rm{d}\Omega} = & -\dfrac{\varepsilon^2 r_e^2 \alpha^2 (m_{A'}^2 - 4\omega^2)^3}{33177600 \pi^2 \omega^4 m^6}(139m_{A'}^4 + 968m_{A'}^2 \omega^2 \notag \\ & + 20016 \omega^4 - 278m_{A'}^4 \cos^2\theta +144 m_{A'}^2 \omega^2 \cos^2\theta \notag \\ & + 13344 \omega^4 \cos^2\theta +  139 m_{A'}^4 \cos^4\theta \notag \\ & -1112 m_{A'}^2 \omega^2 \cos^4\theta + 2224 \omega^4 \cos^4\theta  )
\end{align}
Angular distribution of this reaction is shown in Figure \ref{angulardis}.
Take limit $\varepsilon \rightarrow 1$, $m_{A'} \rightarrow 0$, Eq.(\ref{difsection}) could go back to the case of photon-photon scattering in textbook~\cite{photon-photon}.

\section{Background analysis and constraints}\label{mainpart2}

In the center-of-mass system, kinetic analysis indicates final state photon and dark photon carry energies of $\omega - \dfrac{ m_{A'}^2 }{4 \omega}$ and $\omega + \dfrac{ m_{A'}^2 }{4 \omega}$ respectively.
$\omega = 0.5 $ MeV, $m_{A'} < 1$MeV. Therefore the signal we need to search is monophoton with energy less than $0.5$ MeV plus the missing energy.

As a simplification of the discussion, the detector efficiency is assumed to be 1, which means any final state photon that could travel into the detector can be detected.
Taken into account of shape of the detector, final state photons that can get into the detector must satisfy $|\cos \theta | < 0.8944$~\cite{EPJC}, where $\theta$ is the angle between directions of outgoing photon and incoming photon.

In a $\gamma$-$\gamma$ collider, following processes may happen~\cite{EPJC}:

\noindent $\bullet$ \  Scattering of light-by-light: $\gamma \gamma \rightarrow \gamma \gamma$.
\\ \noindent $\bullet$ \  Breit-Wheeler process as well as cases with final state radiations: $\gamma \gamma \rightarrow e^+ e^-$, $\gamma \gamma \rightarrow e^+ e^- \gamma $, $\gamma \gamma \rightarrow e^+ e^- \gamma \gamma$.
\\ \noindent $\bullet$ \  Compton scattering of a Compton photon and a beam electron: $e^- \gamma \rightarrow e^- \gamma$.
\\ \noindent $\bullet$ \  M\"{o}ller scattering of beam electrons: $e^- e^- \rightarrow e^- e^-$.

The initial photon energy is set to be $0.5$ MeV, which means under this energy scale, electron positron pair could only appear as virtual states, therefore the Breit-Wheeler processes are out of our consideration. On the other hand, beam electron has energy about $200$ MeV, outgoing photons of this ultra-relativistic Compton scattering has very small $\theta$, so most of Compton photons could not travel into the detector. Hence this process is not taken into consideration as well.

\begin{figure}
  \centering
  \includegraphics[width=0.45\textwidth]{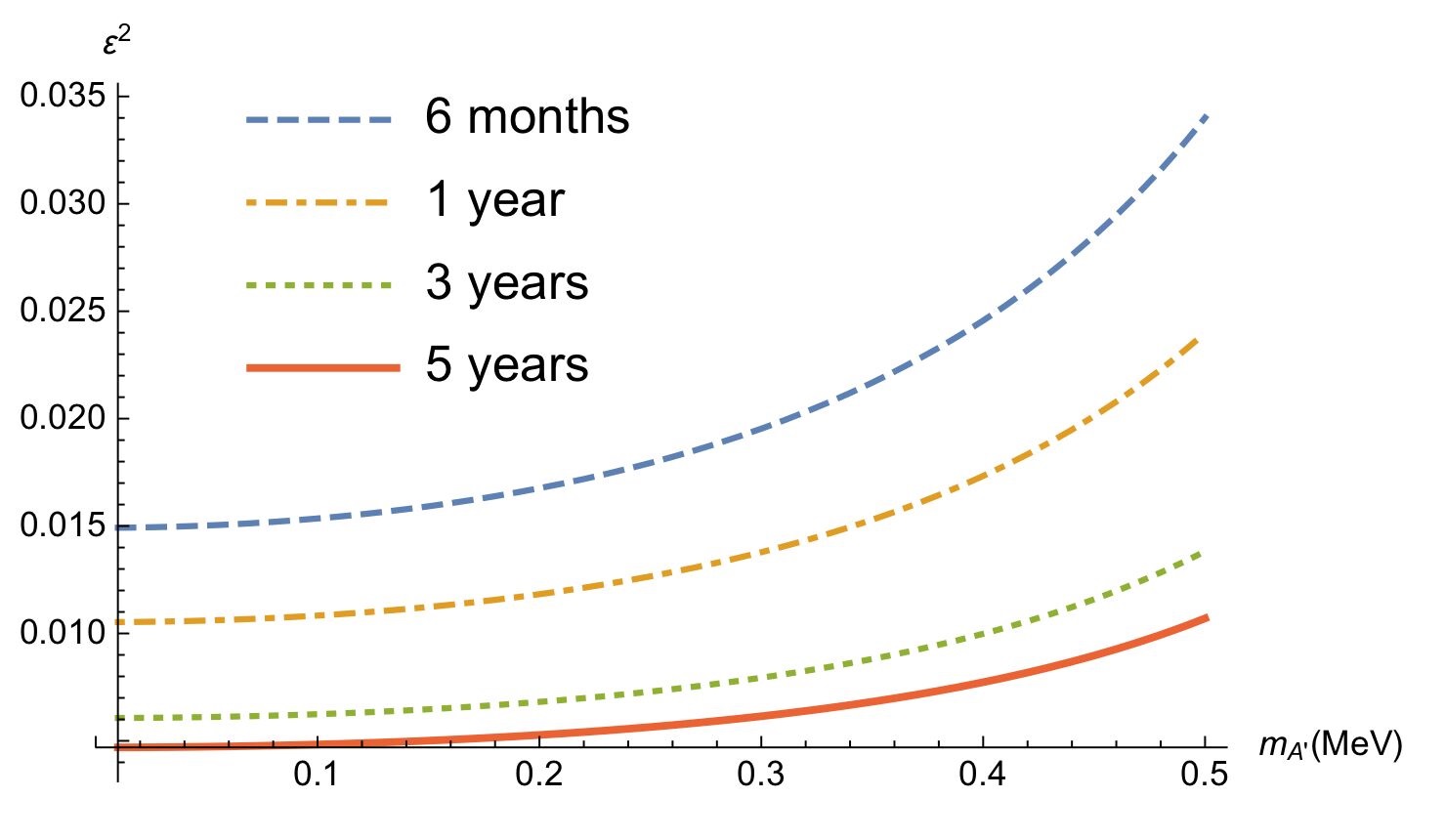}\\
  \caption{Limits on dark photon from $\gamma$-$\gamma$ collider. Orange solid line is constraint with $\gamma$-$\gamma$ collider has a run-time of five years. The other three, blue dashed line is considered under six-months' run-time, brown dot dashed line is of one-year operation and green dot line is three years. }\label{constraints}
\end{figure}

Therefore only scattering of light-by-light is considered as the background. $\dfrac{S}{\sqrt{S+B}} > 1.64 $ is set to give the $95\%$ C.L constraint. Here $S$ and $B$ are signal and background events respectively. The luminosity of $\gamma$-$\gamma$ collider is taken as $L = 4 \times 10^{27} \mathrm{cm}^{-2} \mathrm{s}^{-1}$~\cite{EPJC}. Four time scales are considered: six months, one year, three years and five years. Constraints for each time interval are given in Figure \ref{constraints}, in which orange solid line is limit with five-years' run-time of $\gamma$-$\gamma$ collider. The other three, green dot line, brown dot dashed line and blue dashed line are considered under three-years', one-year' and six-months' collider run-time respectively. Obviously we can see that longer time scale gives more stringent constraint on dark photon.

Current limits on massive dark photon for $m_{A'} < 1$ MeV mainly come from cosmology, astrophysics, atomic experiments, XENON10, TEXONO, etc. COBE/FIRAS gives constraints on dark photon by measuring the distortions of the CMB spectrum, kinetic mixing parameter $\varepsilon \geq 10^{-4}$ are excluded  in mass window $10^{-15} \sim 10^{-11}$ eV ~\cite{cont.1.1}. Constraints on solar hidden photon are given by CAST~\cite{cont.2.1} and SHIPS~\cite{cont.2.2}. There are also limits from solar lifetime(SUN-T, SUN-L), red giants(RG) and horizontal branches(HB)~\cite{cont.2.3}. In atomic expriments, dark photons are constrained by the sensitive test of Coulomb's law on atomic length scales~\cite{cont.3.1}.

\section{Summary}\label{summary}

In this work, the feasibility of searching dark photon in $\gamma$-$\gamma$ collider is considered. With PACKAGE-X, differential cross section of process $\gamma \gamma \rightarrow \gamma A'$ is calculated. For the simplicity of discussion, any final state photon with $|\cos \theta| < 0.8944 $ are assumed to be detected. Taken scattering of light-by-light as background, constraints are given with different time intervals of six months, one year, three years and five years, as shown in Figure \ref{constraints}.

Compared with other constraints on dark photon lighter than $1$ MeV, the upper limit of dark photon in a $\gamma$-$\gamma$ collider is not as stringent as others. This is probably restricted to the capability of the machine, which could not offer higher luminosity under current level of technology. We can expect as energy and luminosity increase in a future $\gamma$-$\gamma$ collider, it will be able to detect a wider dark photon mass window and give better limits.

\acknowledgments

I would like to thank C.Y.Gao, X.Guan, Y.D.Liu and H.Zhang for useful discussions. This work is supported in part by National Natural Science Foundation of China (11655003); Innovation Project of IHEP (542017IHEPZZBS11820, 542018IHEPZZBS12427); the CAS Center for Excellence in Particle Physics (CCEPP).

%%%%%%%%%%%%%%%%%%%%BIBLIOGRAPHY%%%%%%%%%%%%%%%%%%%%%%%%%%%%%%%%%%
%\bibliographystyle{ieeetr}
%\bibliography{DarkPhoton}

\end{document}